\documentclass[useAMS,usenatbib]{mn2e}
\usepackage{graphicx}
\usepackage{bm}

\title[Direct imaging properties of an optical fibered long baseline interferometer]{Optimization of the direct imaging properties of an optical fibered long baseline interferometer}
\author[F.~Patru, D.~Mourard, O.~Lardi\`ere and S.~Lagarde]{F.~Patru$^{1}$\thanks{E-mail:
fabien.patru@obs-azur.fr}, D.~Mourard$^{1}$, O.~Lardi\`ere$^{2}$ \& S.~Lagarde$^{1}$\\
$^{1}$Observatoire de la C\^ote d'Azur, Dpt. Gemini, UMR CNRS 6203, avenue Copernic, 06130 Grasse, France\\
$^{2}$Coll\`ege de France, Observatoire de Haute Provence, 04870 Saint Michel
l'Observatoire, France}
\begin{document}

\date{Accepted . Received ; in original form }

\pagerange{\pageref{firstpage}--\pageref{lastpage}} \pubyear{2006}

\maketitle

\label{firstpage}

\begin{abstract}
Long baseline interferometry is now a mature technique in the
optical domain. Current interferometers are however highly limited
in number of sub apertures and concepts are being developed for
future generations of very large optical arrays and especially with the goal of direct imaging. In this paper, we study the effects of introducing single-mode fibers in direct imaging optical interferometers. We show how the
flexibility of optical fibers is well adapted to the pupil
densification scheme. We study the effects of the truncation
of the gaussian beams in the imaging process, either in the Fizeau mode or in the
densified pupil mode or in the densified image mode. Finally, in the pupil densification configuration, we identify an optimum of the diaphragm width. This optimum maximizes the on-axis irradiance and corresponds to a trade-off between the loss of transmission and the efficiency of the densification.
\end{abstract}


\section{INTRODUCTION}

\citet{lab96} has described the possibility of making direct
snapshot images with interferometric arrays, leading to high
dynamic imaging properties well suitable for stellar surface
imaging and also for coronagraphy for exoplanets finding and
analyzing. Direct imaging can be achieved by the Fizeau
combination \citep{fiz68}, by the pupil densification
corresponding to the hypertelescope concept \citep{lab96}, or by
the image densification proposed in the IRAN concept
\citep{vak04}. In their ground version, these combination schemes require the rearrangement in real time of the projected pupil on the sky and an efficient
cophasing system before the beams are combined. Due to their
flexibility and their spatial filtering properties, single-mode
optical fibers are the right technological choice for these
applications.

In this paper, we briefly present the beam combination schemes for direct imaging. Then, we study the influence of the gaussian field of the fibers on the imaging process. The distribution of intensity in the interferometric image is theoretically written for such imaging systems. Numerical simulations are achieved and criteria are defined to evaluate the performances of pupil densification systems.

\section[]{PRINCIPLES OF BEAM COMBINATION FOR DIRECT IMAGING}

The Fizeau mode preserves the pupil shape by an homothetic mapping
of the entrance pupil to the exit pupil. The Fizeau image of a
point-like source is a fringe pattern, which results from the
multiplication of the impulse response of an aperture by the interferometric pattern
(i.e. the Fourier transform of the Dirac distribution corresponding to
the position of each sub-aperture). If the aperture is highly
diluted (small diameter of sub-pupils and long baselines), the
flux is dispersed in a large diffractive envelope. Only the central peak, which contains little energy, is of interest to interpret the direct image. Post-processing data analysis, like aperture synthesis or specific deconvolution
algorithms, allow to exploit all the collected energy. However, in practice, the
sensitivity is mainly limited by the read-out noise of the instrument.

\begin{figure}
\centering
\includegraphics[width=75mm]{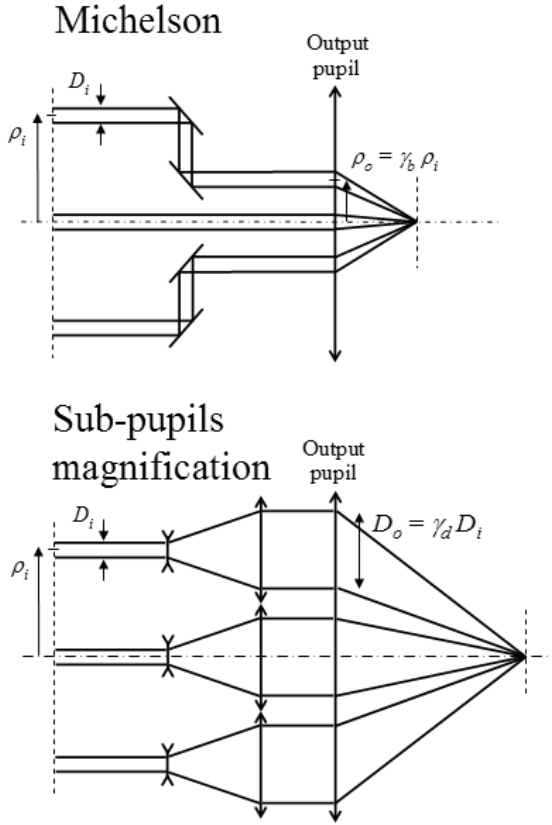}
\caption{Pupil densification (DP) optical schemes using classical optics. A Michelson scheme (up) or an array of inversed Galilean telescopes (down) give the same densified image. For a large array, a combination of both schemes is achieved to bring closer the beams and to adjust their diameters.}
\label{fig:DP scheme}
\end{figure}

\begin{figure}
\centering
\includegraphics[width=75mm]{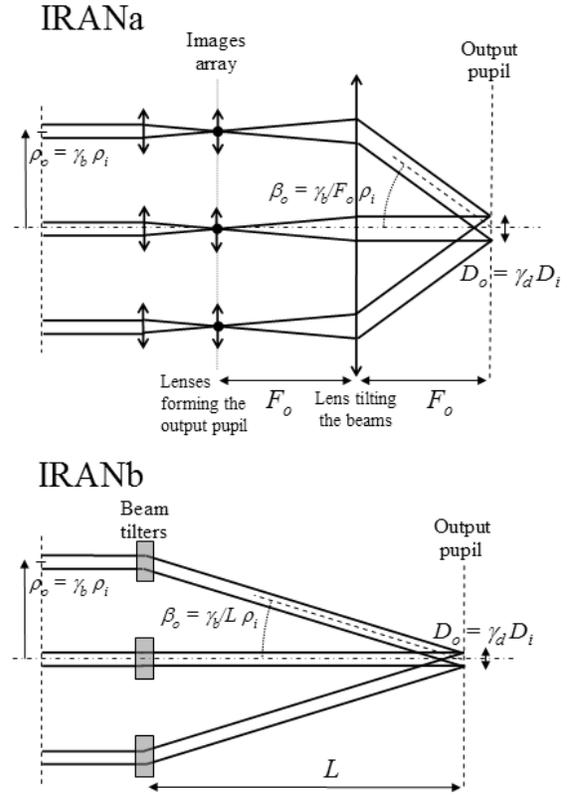}
\caption{Image densification (IRAN) optical schemes using classical optics. IRANa (up) or IRANb (down) are strictly equivalent. The beam tilter in IRANb is composed for example of two tilted mirrors.}
\label{fig:IRAN scheme}
\end{figure}

The pupil densification (DP) modifies the entrance pupil in a densified
exit pupil by increasing the size of the sub-apertures with
respect to their relative distances. This non pure homothetic
transformation corresponds to a "conformal" Michelson scheme. It is performed by moving the positions of the sub-pupils centers nearer without changing the overall pattern. It can also be achieved by magnifying each sub-pupil with an array of inversed Galilean
telescopes, reflective or refractive (Fig. \ref{fig:DP scheme}). Assuming that all apertures have the same diameter, the densification factor is defined as : $\gamma=\gamma_d/\gamma_b$. $\gamma_d=D_o/D_i$ is the scaling ratio of the output and input diameters of each beam, and $\gamma_b=\rho_o/\rho_i$ is the scaling ratio of the output and input radial coordinates of the centers of each sub-pupil. The maximum densification factor is limited when two output densified sub-pupils become tangent.

The characteristics of the interferometric image
depend on the geometry of the exit pupil plane and of the
wavefront errors in the entrance pupil plane. If the exit
sub-aperture diameter increases then the width of the PSF
envelope decreases. Thus, compared to the Fizeau mode, the
hypertelescope mode provides images that are highly luminous.
Indeed, most of the light is concentrated in the central peak surrounded by a halo composed of residual side-lobes. The main interest is the intensification
of the central peak by the factor $\gamma^2$. The diameter of the
usable field, called the direct imaging field (DIF), is equal to
$\lambda/((\gamma-1) D_i)$, where $\lambda$ is the wavelength \citep{lar06}. There is now a "pseudo-convolution" between the object and the image \citep{lab96}.

The image densification (IRAN) consists in forming the interference pattern of all beams in a common output pupil plane. This can be achieved in two ways (Fig. \ref{fig:IRAN scheme}). In the first scheme, called IRANa \citep{vak04}, the sub-images of each aperture are arranged in the intermediate focal plane to reproduce the array configuration at a reduced scale. The fringe pattern is recorded in a common pupil plane relayed by field lenses. In the second scheme, called IRANb \citep{ari04}, each afocal beam is tilted. A relay lens introduced in each beam forms an image of the pupil at the intersection of the beams, where the interferogram is recorded. Both schemes are strictly equivalent. An on-axis star produces a central bright spot at the center of the conjugate stacked pupils.

\begin{figure*}
\centering
\includegraphics[width=15cm]{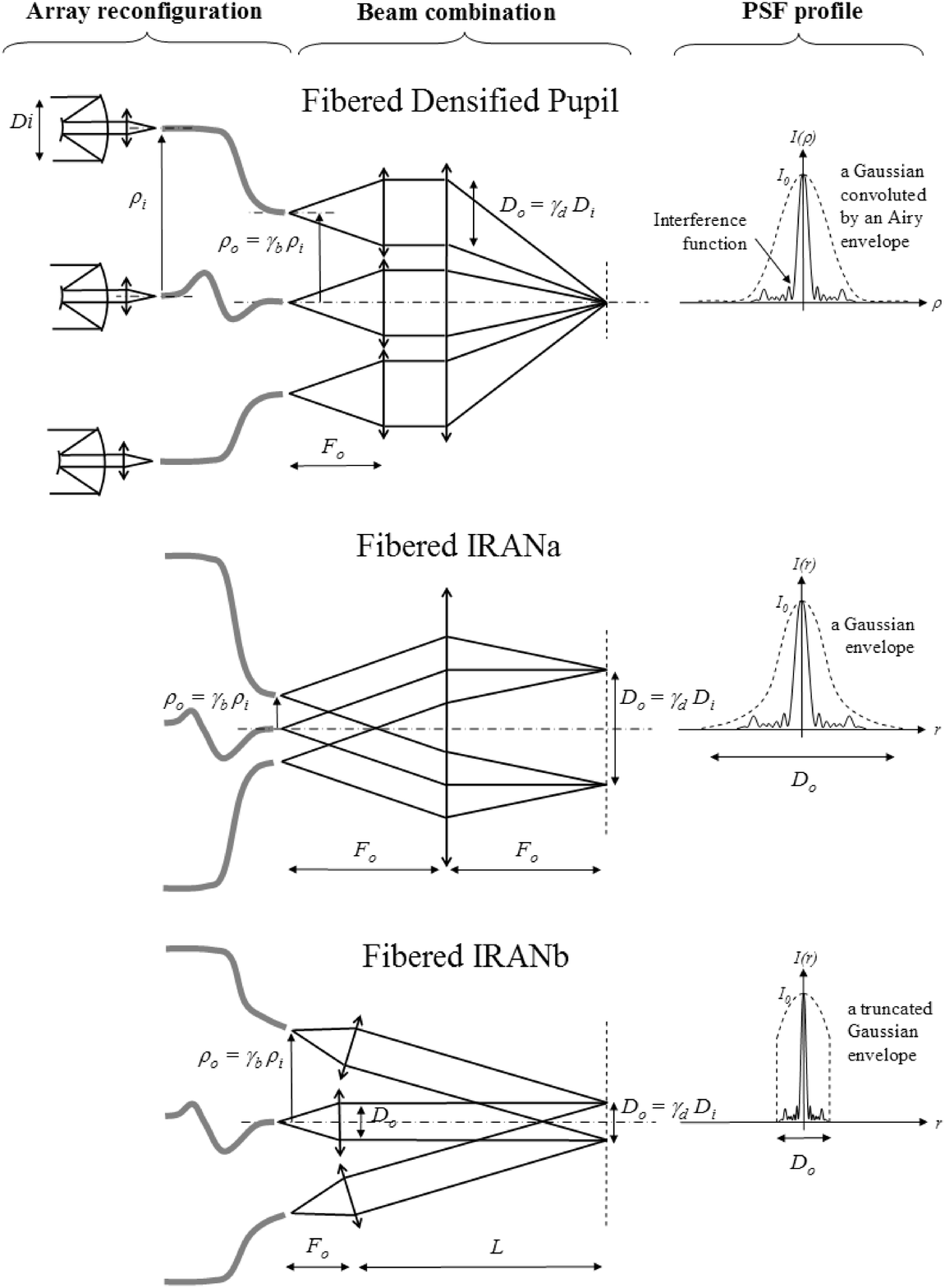}
\caption{Direct imaging optical schemes using single-mode optical fibers. 
For the DP scheme, the positions of the sub-pupils centers are moved nearer or the diameter of the sub-pupils is increased by using the divergence of the beam at the exit of the fiber.
For the IRANa scheme, the images of each sub-apertures are directly defined by the output core of each fiber so that only a back field lens is required. For the IRANb scheme, each beam is collimated by a lens and is easily tilted with the fiber.
However, a high level of densification requires, for both IRAN cases, a very dense remapping of the fiber cores.}
\label{fig:fiber scheme}
\end{figure*}

\section[]{FIBERED VERSIONS OF BEAM COMBINATION SCHEMES}

Single-mode optical fibers appear as a flexible and compact solution
for beam combination and densification. Firstly, the fibers carry the beams from each telescope Coud\'{e} focus up to the combination system (\citealt{per02}, 2006). Secondly, a fiber densifier carries the beams from the entrance to
the exit pupil with the appropriate real-time rearrangement of
the sub-apertures (Fig. \ref{fig:fiber scheme}). Finally, the spatial filtering properties of single-mode fibers are used to properly convert random atmospheric perturbations in the entrance fields in well-controlled photometric fluctuations. It becomes then very easy to reach high accuracy measurements, as has already been demonstrated on the sky with 2 telescopes by IOTA/FLUOR \citep{cou97} and VLTI/VINCI \citep{ker02}, and also in the laboratory with 3 beams for aperture synthesis \citep{del00}.

\subsection{General characteristics of a single-mode fiber}

The properties of the fiber transform the uniform field into a gaussian field. The goal is to properly fit the unlimited gaussian beam to define an output sub-pupil disc. Indeed, the efficiency of the densification is affected by the gaussian
irradiance distribution on each of the output sub-pupils, compared
to a uniform irradiance distribution.

The fundamental mode field distribution for a single-mode fiber is
approximated by a gaussian function. The amplitude distribution at
the exit of the fiber in a plane transverse to the direction of
propagation can be written as \citep{sal91}

\begin{equation}
\psi(r,z)=A_0\frac{\omega_0}{\omega(z)}e^{\frac{-r^2}{{\omega(z)}^2}}
\end{equation}

with $r$ the radial coordinate. $\omega$ is the radius corresponding to the radial distance at which the amplitude of the beam is equal
to $1/e$ time the on-axis amplitude $A_0$. $\omega$ depends on the propagation distance $z$ from the waist of the fiber head

\begin{equation} \label{eq:radius}
\omega(z)=\omega_0\sqrt{1+{\left(\frac{\lambda \,z}{\pi\, \omega_0^2}\right)}^2}
\approx \frac{\lambda \,z}{\pi\, \omega_0}
\end{equation}

with $\omega_0$ the radius of the waist. The approximation is valid far from the core of the fiber ($z>>1$). The radius $\omega_0$ depends on the characteristics of the fiber and is described by the empiric formula

\begin{equation}
\omega_0=a \left(0.65+\frac{1.619}{V^{1.5}}+\frac{2.879}{V^6} \right)
\end{equation}

where $a$ is the radius of the fiber core. V is the normalised
cutoff frequency of the fiber defined as

\begin{equation}
V=2\pi \,N_A \,a/{\lambda}
\end{equation}

with $N_A$ the numerical aperture of the fiber. $V$ is less then
$2.405$ for a single-mode fiber.

We can then write the irradiance distribution at the output of the
fiber:

\begin{equation}
I(r,z)={|\psi(r,z)|}^2=I_0{\left(\frac{\omega_0}{\omega(z)}\right)}^2 e^{\frac{-2r^2}{{\omega(z)}^2}}
\end{equation}

where $I_0={A_0}^2$ is the on-axis irradiance.

The total power $P_i$ incident on the pupil plane is given by
\begin{equation} \label{eq:p incident}
P_i=2\pi \int_0^{+\infty} I(r,z) \,r \,dr = \frac{\pi \,I_0 \,{\omega_0}^2}{2}
\end{equation}

\subsection{Influence of a truncated gaussian beam}

The case of a truncated gaussian beams has already been studied for one aperture (\citealt{buc67}, \citealt{mah86}, \citealt{nou01}).
We consider here that the image is formed by an aberration-free
spherical lens of focal length $F_o$ and diameter $D_0=2 R_0$. The
gaussian beam is then diaphragmed by the circular aperture of the
lens. We defined the normalized beam radius $k$ as the ratio:

\begin{equation}
k=\frac{R_o}{\omega(F_o)}
\end{equation}

By using Eq. \ref{eq:radius}, $k$ can be expressed as a function of the output beam aperture $R_o/F_o$ :

\begin{equation}
k=\frac{\pi\, \omega_0}{\lambda} \,\frac{R_o}{F_o}
\end{equation}

The truncated gaussian amplitude distribution of a sub-pupil can be written as

\begin{equation} \label{eq:amplitude 1}
\psi(r,k)=A_0 \,\frac{\omega_0}{R_o} \, k
\,e^{\frac{-k^2}{{R_o}^2}\,r^2} \,\Pi_{R_o}(r)
\end{equation}

with $\Pi_{R_o}(r)=1$ if $r\leq R_0$ and $\Pi_{R_o}(r)=0$ otherwise.

Refering to \citet{has79}, for an incident power $P_i$ and for a lens with an aperture of $F_o/D_o$, the focal-plane irradiance is given by 

\begin{eqnarray} \label{eq:focal irrad 1}
I(\rho,k) = {|\mathrm{FT}(\psi(r,k))|}^2 = \frac{8\,P_i}{\pi\,\omega_0^2} \left(\frac{k}{R_o}\right)^4 \nonumber\\
{\left|\int_0^{R_o} \frac{R_o}{k}\, e^{-\left(\frac{k}{R_o}\,r\right)^2} J_0{\left(\frac{2\pi}{\lambda \,F_o} \,r \cdot \rho \right)} r \,dr \right|}^2
\end{eqnarray}

where $r$ and $\rho$ denote a position respectively in the pupil plane and in the
focal plane. FT refers to the Fourier transform.
$J_0$ is the Bessel function of the first kind of order zero.
If we let $\rho=0$, we can write the maximum on-axis irradiance as

\begin{equation}
I(0,k) = \frac{8\,P_i}{\pi\,\omega_0^2} \left(\frac{k}{R_o}\right)^2
{\left|\int_0^{R_o} e^{-\left(\frac{k}{R_o}\,r\right)^2} r \,dr \right|}^2
\end{equation}

\begin{equation}\label{eq:i0_mono}
I(0,k) = \frac{2\,P_i\,R_o^2}{\pi\,\omega_0^2} \frac{{(1-e^{-k^2})}^2}{k^2}
\end{equation}

The maximum on-axis irradiance is obtained by equating to zero its derivative
with respect to $k$. $\frac{{\delta}I(0)}{{\delta}k}=0$ leads to the equation

\begin{equation} \label{eq:deriv irrad}
e^{k^2}=1+2k^2
\end{equation}

The solution of Eq. \ref{eq:deriv irrad} is approximately $k=1.12$ (\citealt{sie89}, \citealt{yur95}).

The total power transmitted by the truncated beam is equal to

\begin{equation}
P_{tot}=2\pi \int_0^{R_o} I(\rho,k) \,r \,dr = \frac{\pi \,I_0 \,{\omega_0}^2}{2}(1-e^{-2k^2})
\end{equation}

Hence the fraction of the power transmitted after truncation of the
sub-pupil corresponds to
\begin{equation} \label{eq:p trans}
P_{transm}=P_{tot}/P_i=1-e^{-2k^2}
\end{equation}

\begin{figure*}
\begin{center}
\begin{tabular}{lcr}
\includegraphics[width=55mm]{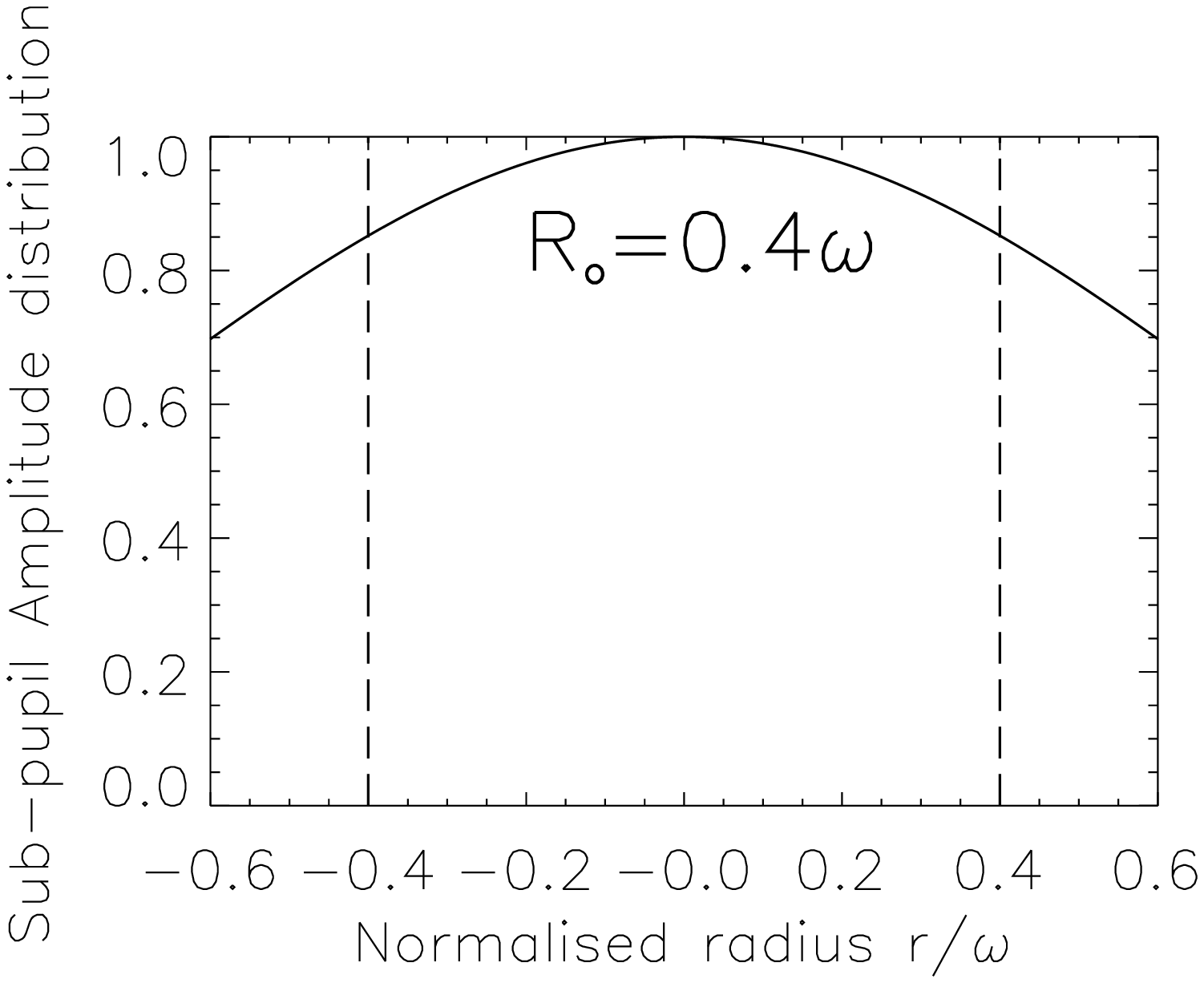} & \includegraphics[width=55mm]{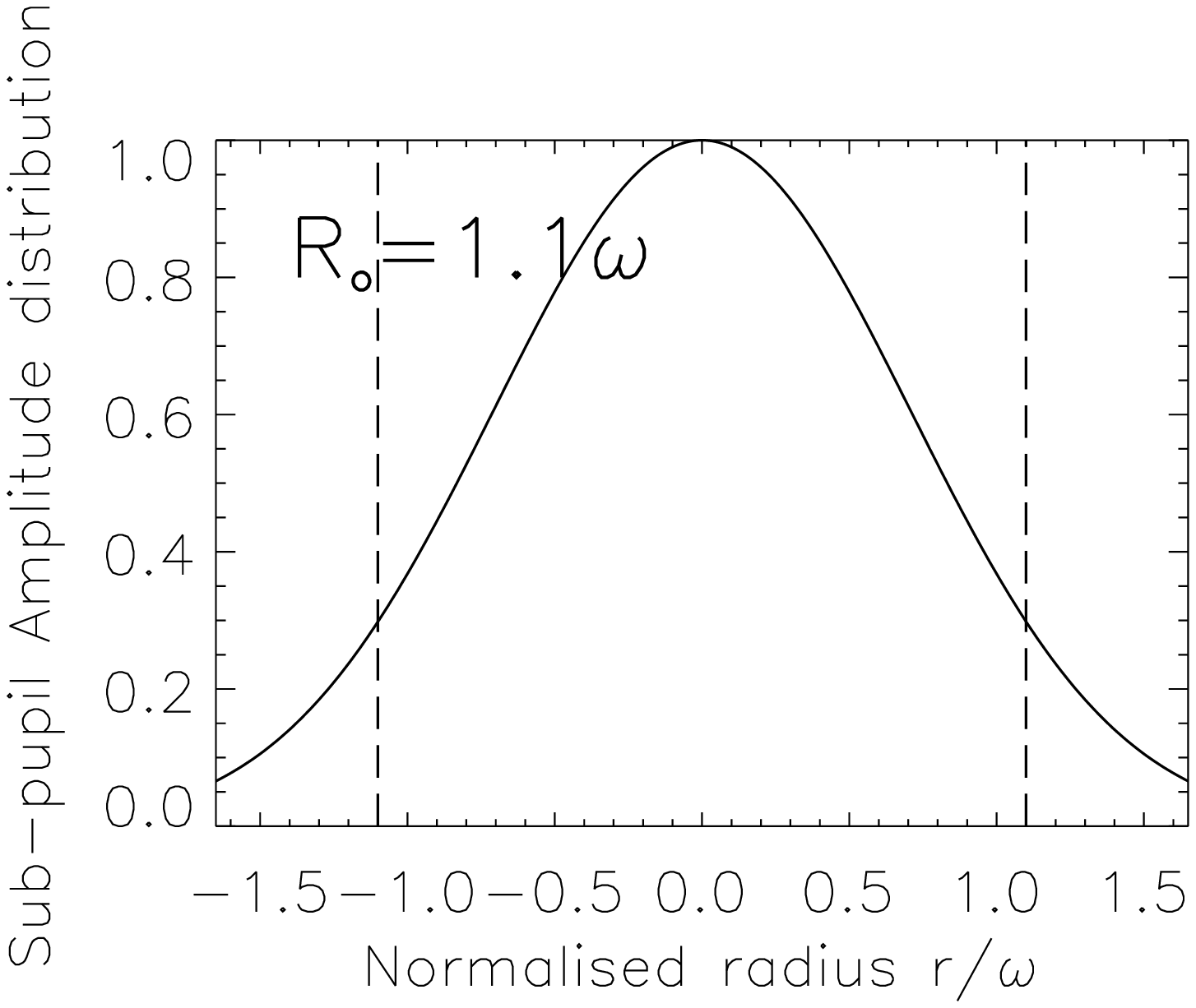} & \includegraphics[width=55mm]{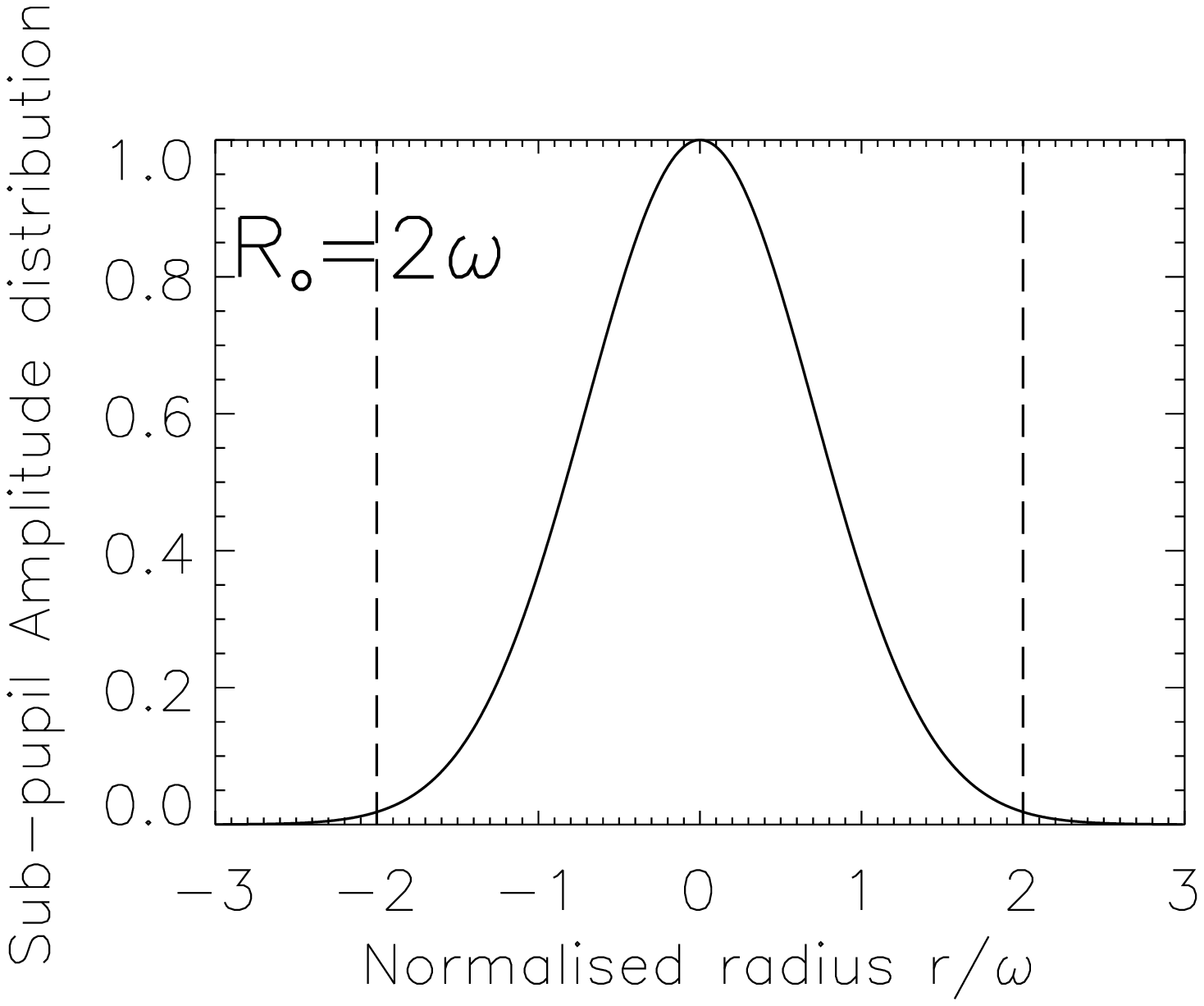} \\
\includegraphics[width=55mm]{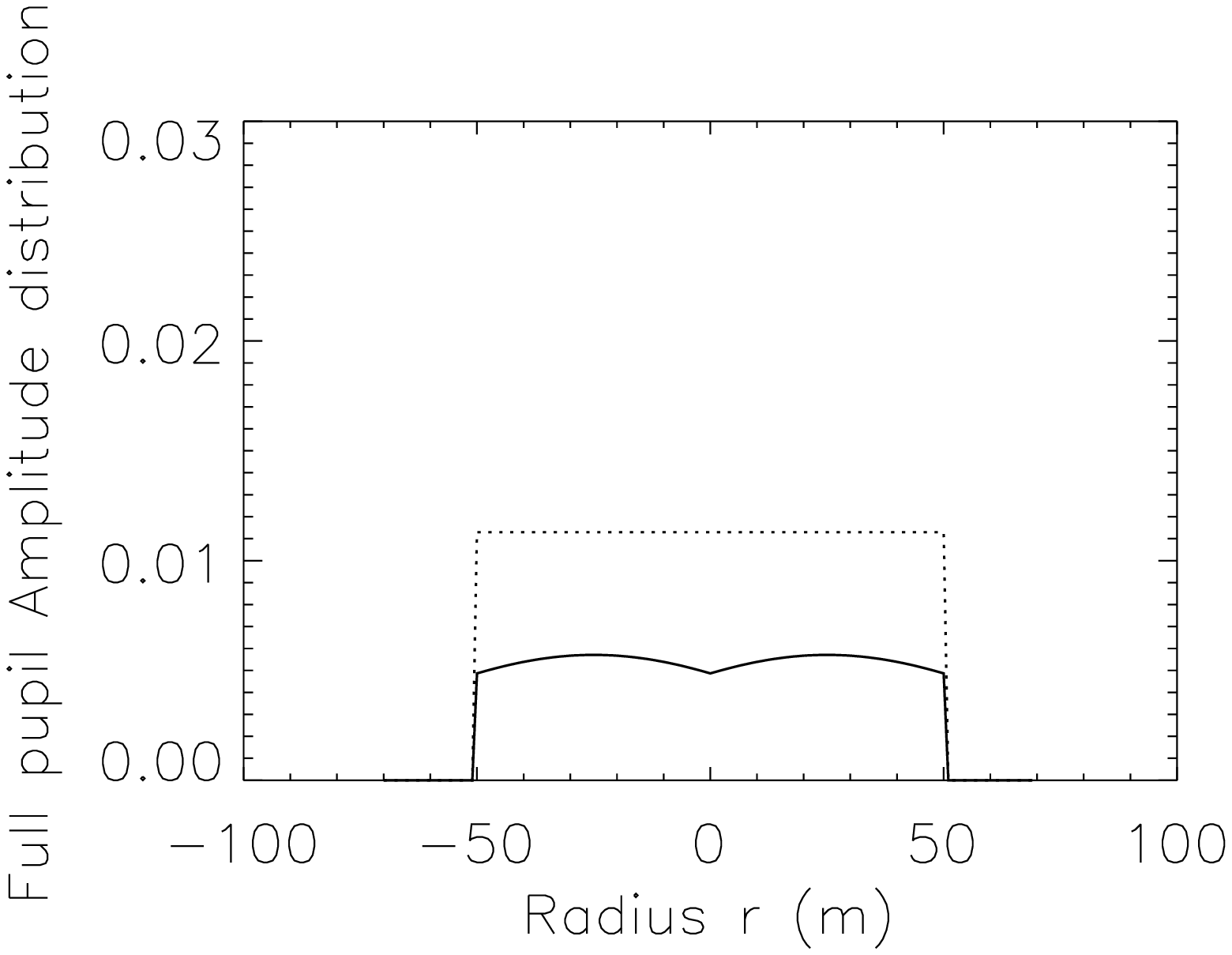} & \includegraphics[width=55mm]{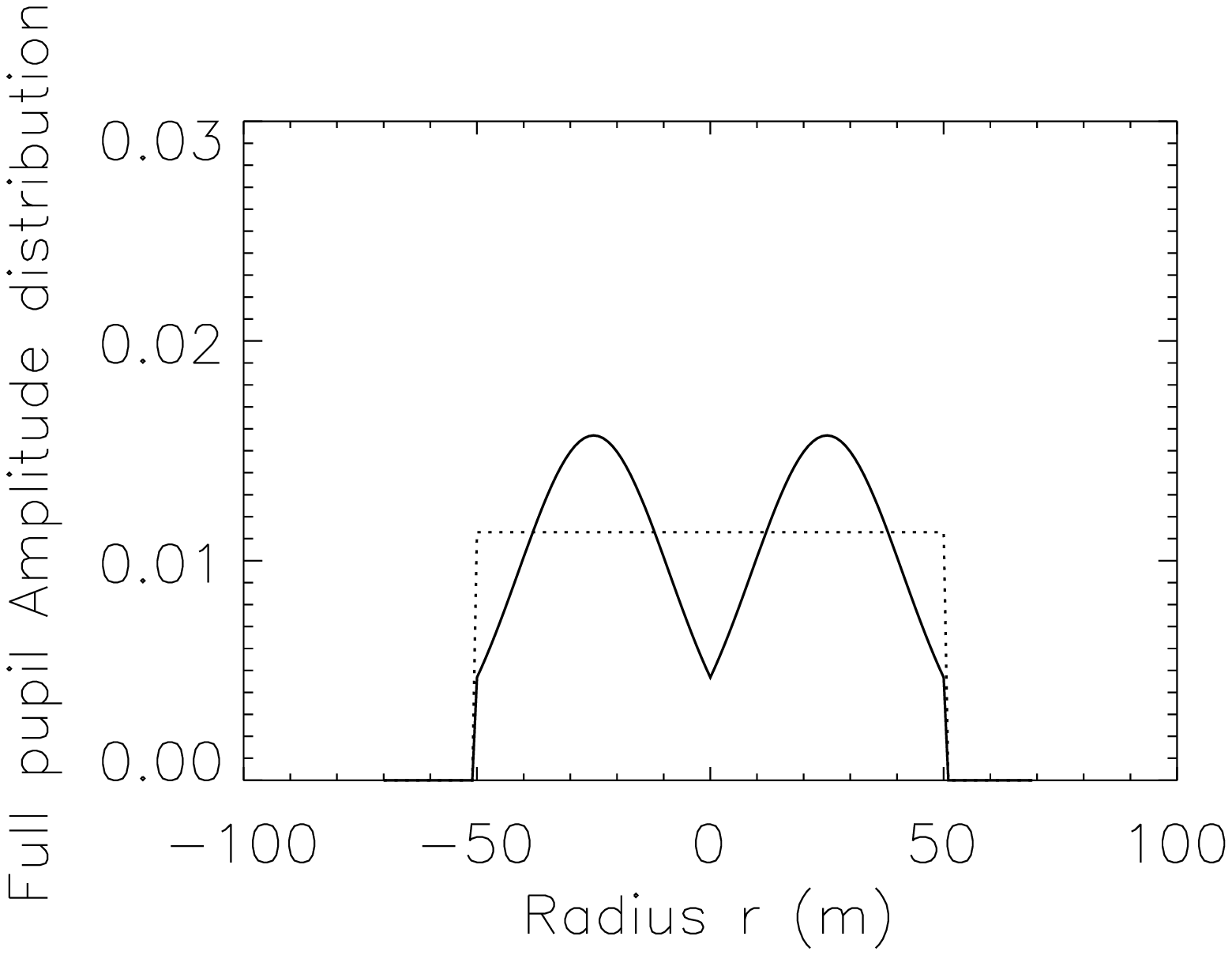} & \includegraphics[width=55mm]{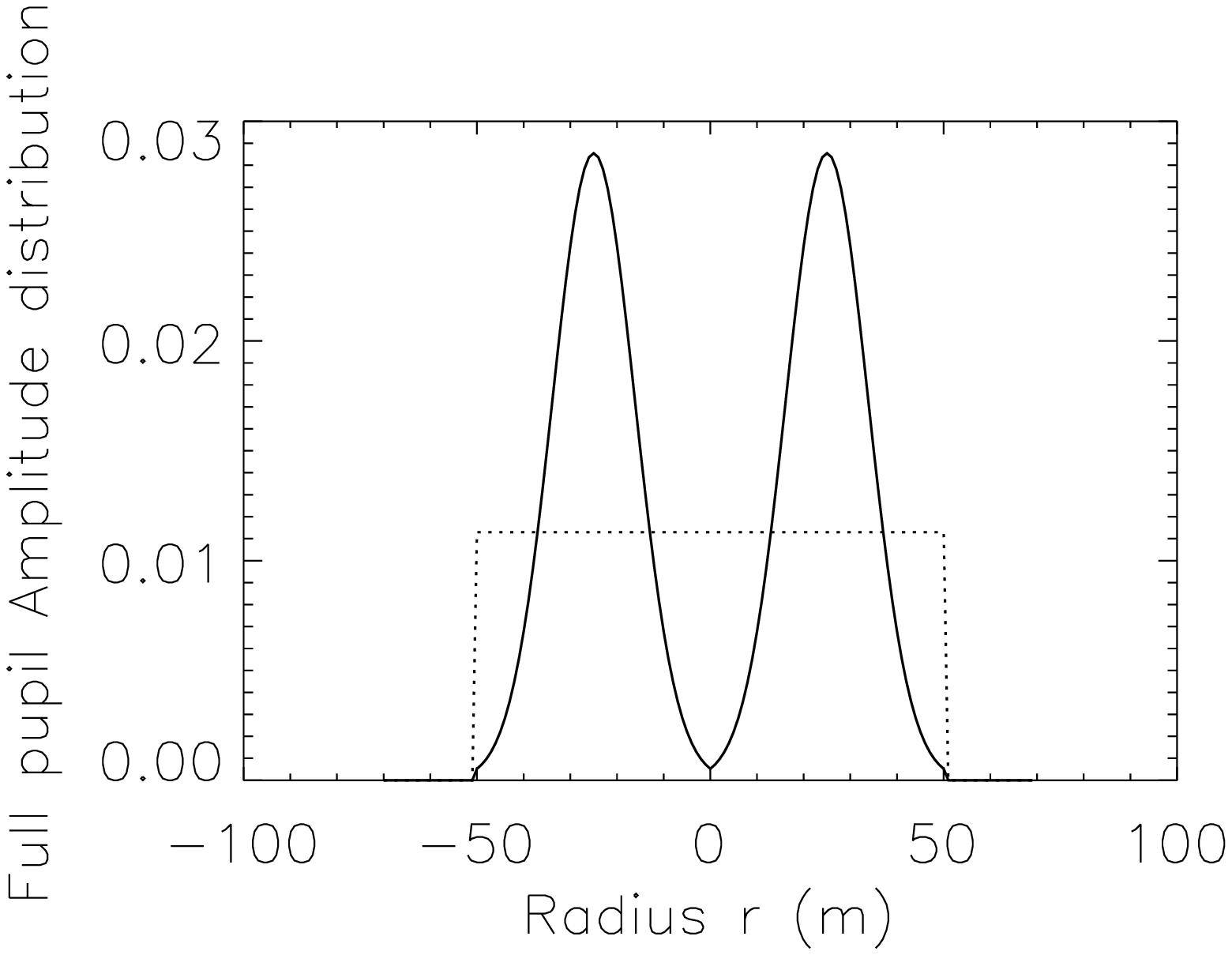} \\
\includegraphics[width=55mm]{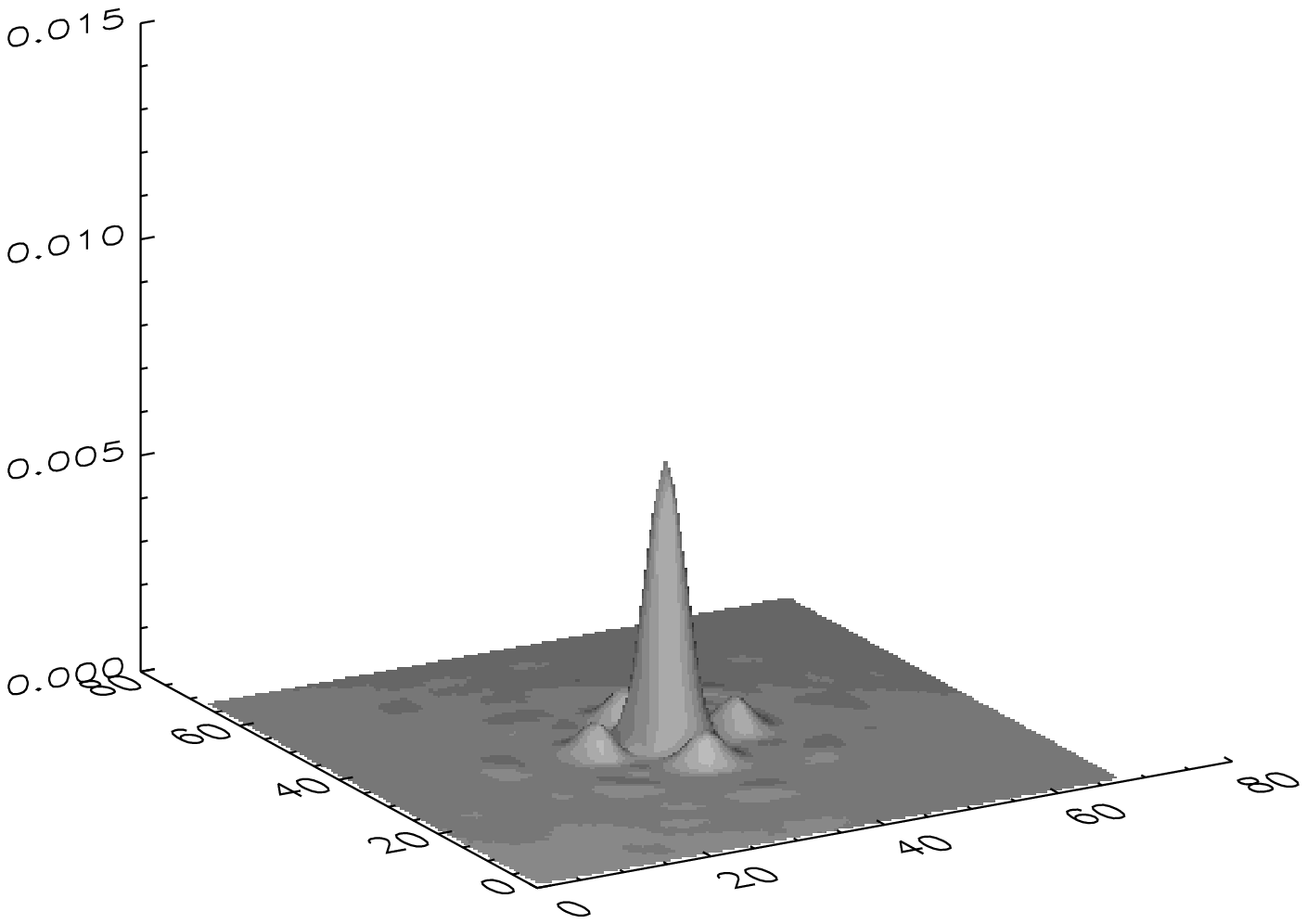} & \includegraphics[width=55mm]{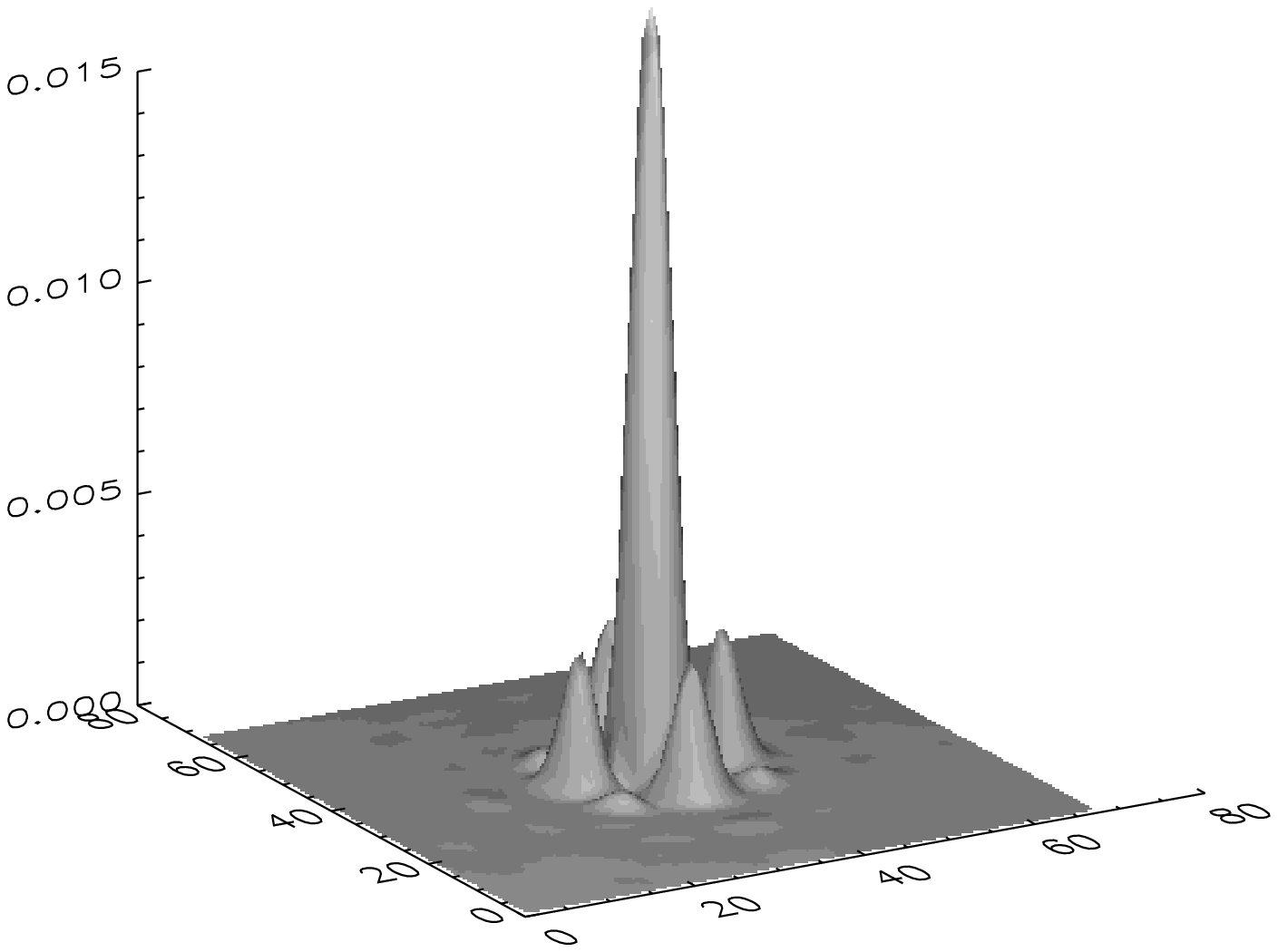} & \includegraphics[width=55mm]{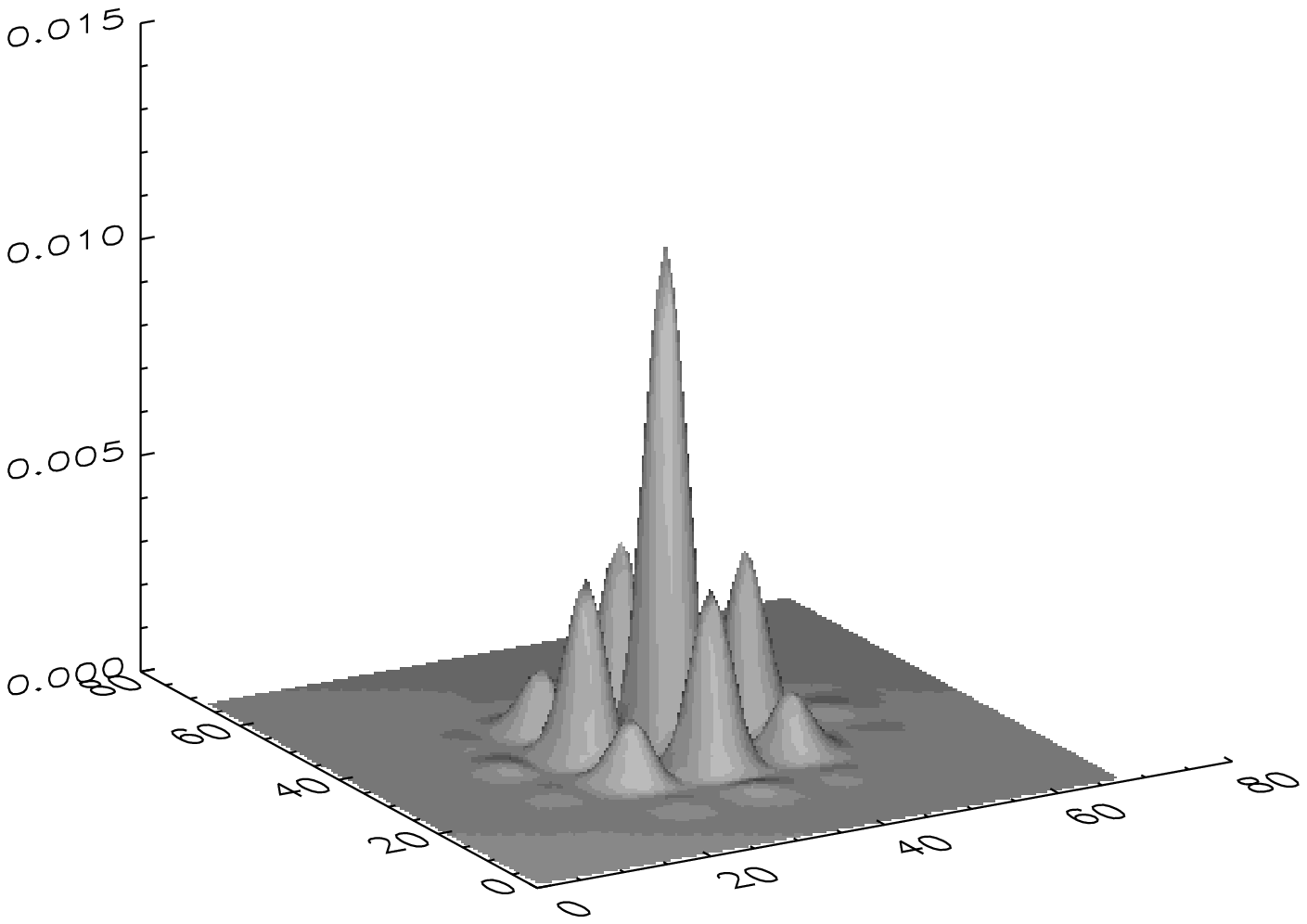} \\
\includegraphics[width=55mm]{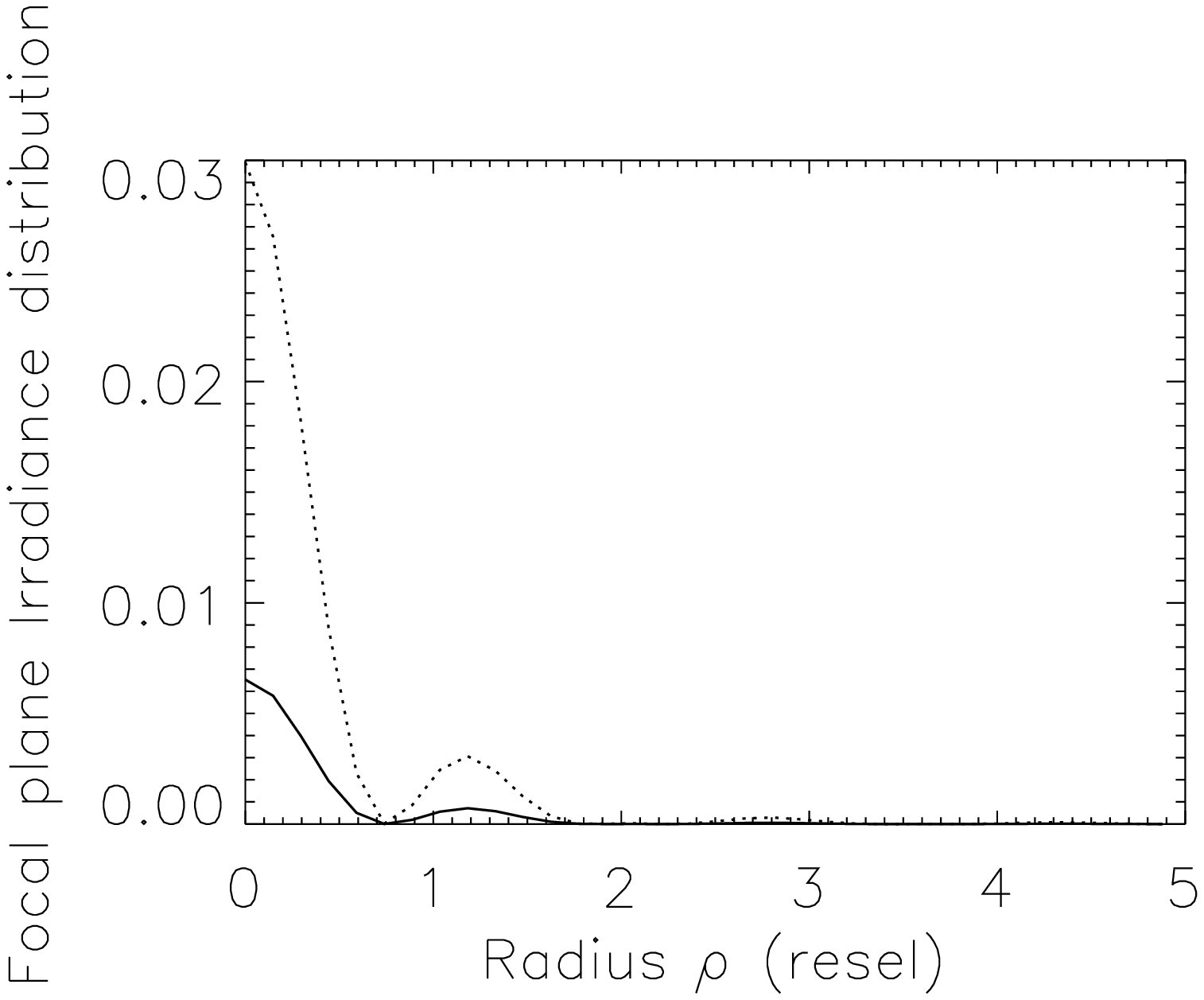} & \includegraphics[width=55mm]{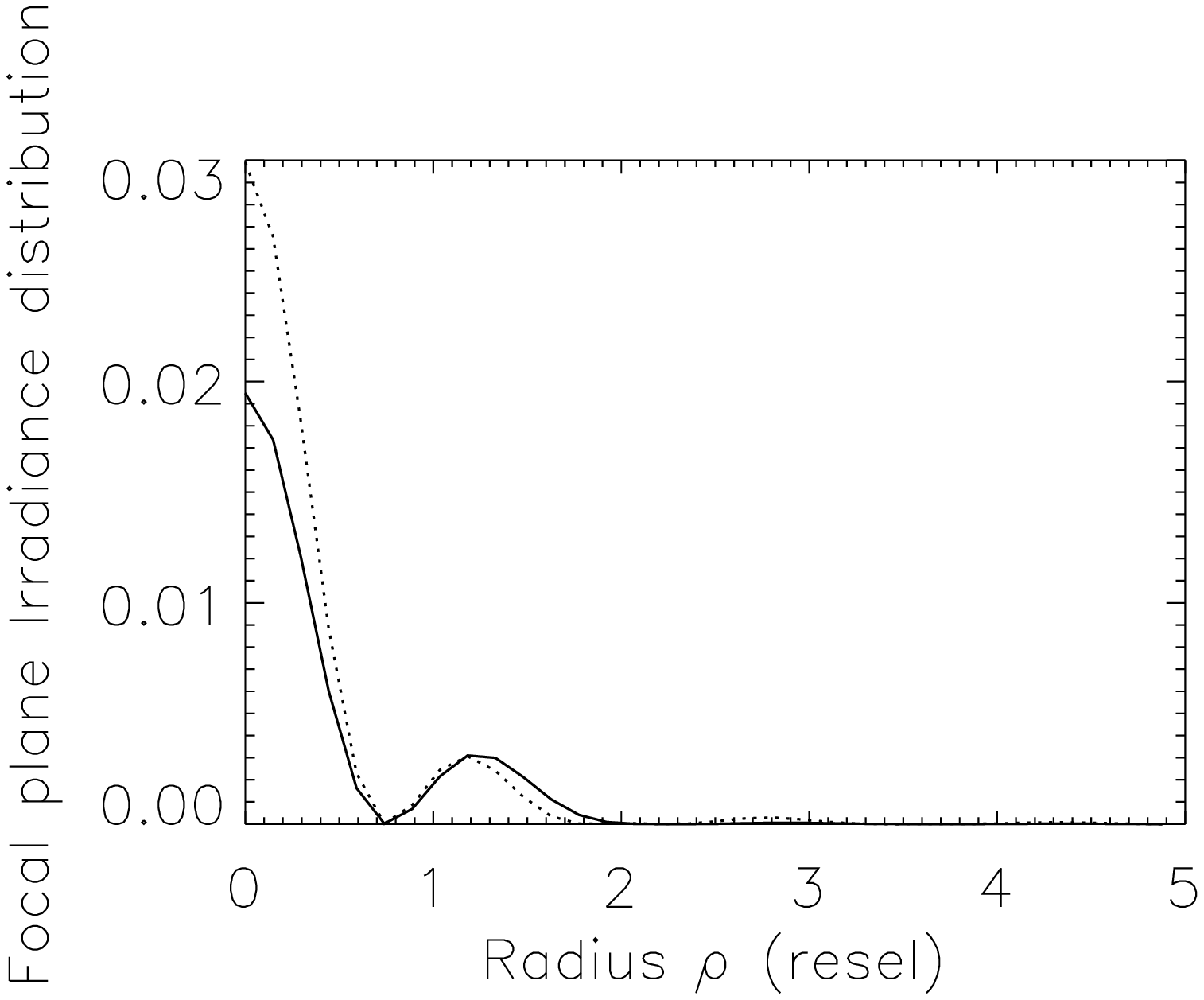} & \includegraphics[width=55mm]{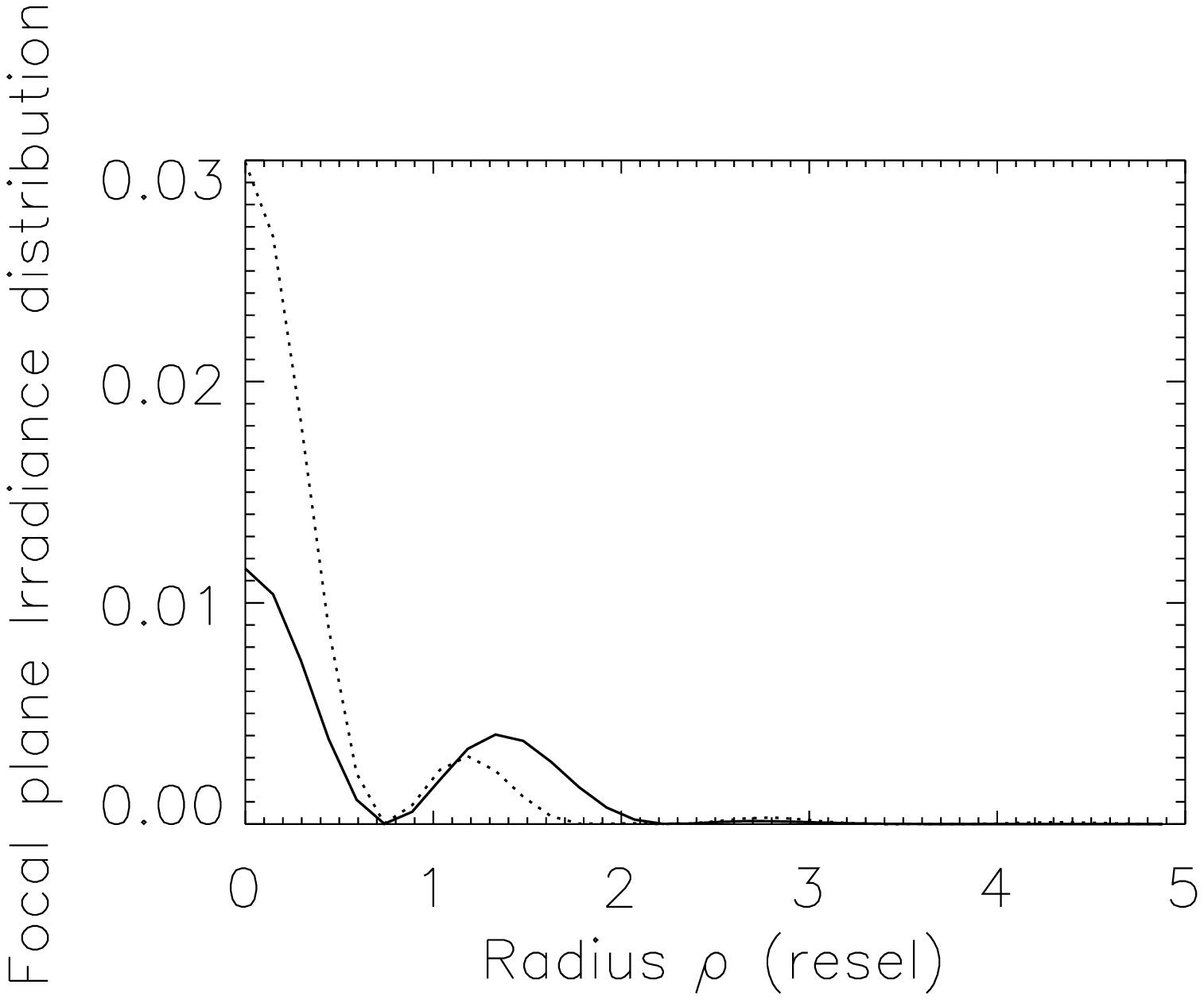} \\
\end{tabular}
\end{center}
\caption{Effects of the gaussian field of the fibers on the imaging process, for various values of the diaphragm radius $R_o$ as a fraction of the beam radius $\omega$. From up to down : Profile of one gaussian beam with the corresponding diaphragm, cross-section of the output pupil of the fiber densifier in the maximum densification case ($\gamma=10$), densified PSF, cross-section of the PSF. The case of the uniform illumination is represented in dotted lines. For strongly truncated sub-pupils (left), the pupil-plane irradiance distribution is almost uniform and the image pattern is narrow, but little power is transmitted. For weakly truncated sub-pupils (right), the power in the focal-plane is rejected from the center to the far peaks, as if the full pupil is diluted. An optimum (middle) provides the maximum on-axis irradiance.}
\label{fig:gam5 vs 3k}
\end{figure*}

\subsection{Application to an interferometric array}

We will now consider the case of an interferometer defined as an
array of $N_{tel}$ sub-pupils. According to Eq. \ref{eq:amplitude 1}, we can write the amplitude distribution of the exit pupil plane as

\begin{equation}
\psi_{array}(r,k) = \psi(r,k) \otimes \sum_{n=1}^{N_{tel}}
\delta(r-\rho_o(n))
\end{equation}

where $\rho_o(n)$ denotes the position of the
sub-pupil of index $n$ in the exit pupil plane. $\otimes$ is the
convolution operator and $\delta$ is the Dirac function.

In a DP combiner, each gaussian beam is truncated and collimated by a lens. The intensity distribution in the image plane is

\begin{equation}
I_{array}(\rho,k)=\left|\,\mathrm{FT}\left(\psi_{array}(r,k)\right)\right|^2
\end{equation}

which can be written as

\begin{eqnarray}
I_{array}(\rho,k) = \left|\,\mathrm{FT}\left(\psi(r,k)\right)\right|^2 \nonumber\\
\times \left|\sum_{n=1}^{N_{tel}}
\exp\left({\frac{-2\,i\,\pi}{\lambda\,F_o} \,\rho \cdot \rho_o(n)}\right)\right|^2
\end{eqnarray}

By replacing $R_o$ by $\gamma\,R_i$ in equation \ref{eq:i0_mono}, we can write the maximum on-axis irradiance as

\begin{equation}
I_{array}(0,k) = I(0,k) = \frac{2\,P_i\,R_i^2}{\pi\,\omega_0^2}\,\gamma^2\, \frac{{(1-e^{-k^2})}^2}{k^2}
\end{equation}

Although the notion of pupil vanishes with fibers, the ouput sub-pupil width is defined by the truncated beam diameter $D_o$ of the diaphragm.
Thus, $\gamma=\gamma_d/\gamma_b$, as defined previously.
$\gamma_b$ corresponds to the magnification factor of the interferometric image in the focal plane. To simplify, we assume that $\gamma_b=1$.

The experimental arrangement allows independent adjustment of the two parameters $\gamma$ and $k$.
The diaphragm radius $R_o$ is chosen in such a way that $R_o=\gamma_d\, R_i$.
The normalized radius $k$ is adjusted by the focal length $F_o$.

It is shown by equation 19 that the maximum on-axis irradiance is independent of the geometry of the array. It depends only on the diffraction envelope of a sub-pupil (a gaussian function convoluted with an Airy function).
Thus, the densification maximizes the on-axis irradiance when $\gamma$ increases whatever $k$. For a given $\gamma$, the maximum on-axis irradiance is obtained with $k=1.12$, as in the case of a single aperture \citep{sie89}.

In an IRAN combiner, the intensity distribution is similar, the difference being only in the envelope shape. Thus, the intensity recorded in the pupil plane is \citep{vak04}

\begin{equation}
I_{array}(r,k) = \left|\,\left(\psi(r,k)\right)\right|^2
\times \left|\sum_{i=1}^{N_{tel}}
\exp\left({\frac{-2\,i\,\pi}{\lambda\,F_o} \,r \cdot \rho_o(n)}\right)\right|^2
\end{equation}

As the maximum of the envelope (a truncated or untruncated gaussian function) is
constant whatever $D_o$, the truncation has no effect on the
central peak intensity and reduces only the field by vignetting. Then, the truncated radius is chosen such that $k>2$ to transmit the total power.

\section[]{PERFORMANCE ANALYSIS OF A DENSIFIED PUPIL FIBER COMBINER}

In the previous section, it was seen that there is no optimization for the diaphragm of the sub-pupils in the case of IRAN. Concerning the DP scheme, numerical simulations have been performed to study the effect of truncated gaussian beams on the focal-plane irradiance distribution. As the main interest is the sensitivity gain, we use criteria aiming at maximizing the irradiance contained in the central interferometric peak, compared to the irradiance in the side-lobes. We show that the optimum truncated radius is a trade-off between the loss of transmission and the efficiency of the densification.

The simulations are performed in a perfect case, without any noise or wavefront errors and for an on-axis star. We consider a pupil composed of 4 circular sub-apertures on a large square pattern with a side length of $50~m$ and a telescope diameter of $D_i=5~m$. Then the maximum densification corresponds to $\gamma=10$. With an operating wavelengh of $\lambda=0.6~{\mu}m$ and a maximum baseline of about $76~m$, the spatial resolution ($resel=\lambda/B_{max}$) reaches $1.6~mas$. The DIF equals to $1.7~resels$.
Each sub-pupil has a gaussian irradiance distribution with a common free parameter $k$.

Fig. \ref{fig:gam5 vs 3k} presents the calculated Point Spread Functions in the maximum densification case for various values of the $k$ parameter.
The intensity is normalized so that the total transmitted power is unit. It means that the integral of the irradiance of the full pupil equals to 1 with a uniform illumination.
In the case of a gaussian illumination of the sub-pupils, the total transmitted power is weighted by $0.8$, which corresponds to the optimum coupling efficiency of a single-mode fiber \citep{sha88}. 

Finally, in the following parts, all the parameters are normalized so that each value of the gaussian illumination is divided by the corresponding value of the uniform illumination ($k=0$).
This normalization allows to compare the effects of the gaussian beams to the uniform illumination case.

\begin{figure}
\centering
\includegraphics[width=70mm]{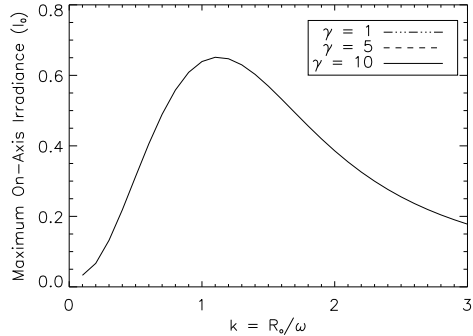}
\caption{Variation of the on-axis irradiance as a function of the truncation ratio $k$ for various values of the densification factor $\gamma$. The 3 curves with $\gamma=\{1,5,10\}$ are superimpozed. Maximizing $I_0$ yields the value of $k=1.12$ whatever $\gamma$. Thus, the densification factor $\gamma$ has no influence of the optimum k value. Note that the 3 cases shown in Fig. \ref{fig:gam5 vs 3k} correspond to $k=0.4$, $k=1.1$ and $k=2$.}
\label{fig:var i0}
\end{figure}

\subsection{On-axis irradiance}

The first criterion consists in maximizing the on-axis irradiance $I_0$. In the case of our 4 telescopes configuration, numerical simulations give the value of $k=1.12$ whatever $\gamma$. The results of the simulation are shown in Fig. \ref{fig:var i0}, where $I_0$ is plotted as a function of $k$.

The criterion can also be achieved by maximizing the encircled energy contained in the main central lobe given by 

\begin{equation} \label{eq:p encircled}
E_0(\theta_0)=2\pi \int_0^{\theta_0} I_{array}(\rho) \,\rho \,d\rho
\end{equation}

where $\theta_0$ corresponds to the first minimum from the center of the focal plane irradiance distribution. This second criteria leads to the same optimum value of the truncation ratio : $k=1.12$.

\subsection{Transmission}

The fractional transmitted power in Eq. \ref{eq:p trans} remains valid whatever the number of sub-apertures. This equation shows that the transmission does not depend on the densification factor $\gamma$. Indeed, the global intensity picked-up by the detector is fixed by the diaphragm radius $k$ of each beams.
The fraction of power is equal to $22$ per cent when $k=0.4$, $73$ per cent when
$k=1.1$, and practically $80$ per cent when $k=2$, corresponding to the maximum coupling
efficiency of a single-mode fiber. Then, for the optimum $k$ value, $20$ per cent of power is lost owing to the fiber and about $8$ per cent owing to the diaphragm.

\subsection{Image quality}

We will now consider some criteria aiming at quantifying the image
quality of the direct image. We have estimated firstly the fraction of energy contained in the central peak and secondly the level of the residual halo.

\begin{figure}
\centering
\includegraphics[width=70mm]{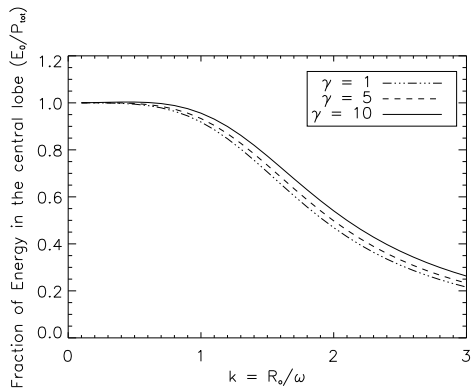}
\caption{Variation of the ratio of the energy of the central lobe
divided by the total energy as a function of $k$ for various values of $\gamma$. The more
the diaphragm radius is increased, the more the light is spread from the central peak to the side-lobes in the halo. This effect is more drastic for an higher level of densification.}
\label{fig:var frac central energy}
\end{figure}

Fig. \ref{fig:var frac central energy} shows the fraction of energy contained in the central peak as a function of $k$, defined as the ratio of central lobe energy
divided by the total energy. This is an estimator of the spread effect of light, which reflects the efficiency of the densification. If the truncation ratio is low, typically $k{\leq}0.5$, this fraction is unaffected and is the
same as in the uniform illumination case. Then it decreases
rapidly up to around $k=2.5$. The shape of the curve is the same for any level of densification and is rather pronounced for $\gamma=10$. For an optimum $k$ radius, the central peak shows a diminution of $11$ per cent, $10$ per cent and $8$ per cent, respectively for $\gamma=1$, $\gamma=5$ and $\gamma=10$, compared to the uniform beams.

\begin{figure}
\centering
\includegraphics[width=70mm]{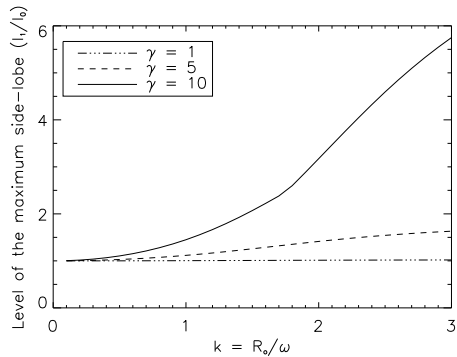}
\caption{Variation of the ratio of the secondary maxima irradiance
divided by the central lobe irradiance as a function of $k$ for various values of $\gamma$.  The more the diaphragm radius is increased, the more the maximum level of the residual side-lobes is high, compared to the central peak. This effect is drastic in the maximum densification case ($\gamma=10$). It has practically no effect in the Fizeau case ($\gamma=1$).}
\label{fig:var ratio}
\end{figure}

In completeness, Fig. \ref{fig:var ratio} represents the ratio of the irradiance of the
secondary maxima divided by the on-axis irradiance $I_0$. This is an estimator of the maximum level of the residual halo surrounding the central peak. This parameter must be improved to limit the confusion problem \citep{ria04}, which corresponds to the crowding effect between the different sources present in the field of view. The source crowding is due to the convolution process of the observed sources with the PSF, as with a classical telescope (single aperture). This effect becomes important when the instrument has a weak quality for its PSF and it is directly related to the PSF extension. That is why the maximum densification scheme is well adapted. Moreover, the use of single-mode fibers requires a suitable diaphragm radius to limit the corruption of the image quality.

\section[]{DISCUSSION AND PERSPECTIVES}

If a fiber image densifier (IRAN) is used, the width of the diaphragm of the output beams is sufficiently enlarged to maximize the transmitted power. Less than $1$ per cent of the flux is lost if

\begin{equation}
\frac{R_o}{F_o} > 2 \, \frac{\lambda}{\pi\, \omega_0}
\end{equation}

However, we have shown that there is an optimum for the diaphragm diameter $D_o$ when a fiber pupil densifier (DP) is used. This optimum is firstly linked to the truncated gaussian irradiance distribution in a sub-pupil, which provides the maximum $I_0$ in the focal plane. The optimum is also decorrelated from the geometry of the array, as demonstrated in section 3. Simulations in section 4 show that it can be interpreted as a trade-off between the loss of transmission and the efficiency of the densification. The optimization aims at maximizing the fractional energy of the central lobe and minimizing the maximum level of the residual halo, given by the highest side-lobe. The main parameters corresponding to Fig. \ref{fig:gam5 vs 3k} are resumed in table \ref{tab:densif vs k} in the maximum densification case.

\begin{table}
\begin{center}
\begin{tabular}{lcccc}
\hline							
$k$	&	0.4	&	1.1	&	2	\\
\hline							
$\psi(R_0)/\psi(0)$	&	0.85	&	0.30	&	0.02	\\
$I_0$	&	0.22	&	0.65	&	0.39	\\
$E_0$	&	0.22	&	0.67	&	0.42	\\
$I_1/I_0$	&	1.07	&	1.55	&	3.42	\\
$E_0/P_{tot}$	&	1.00	&	0.92	&	0.52	\\
$P_{transm}$	&	0.22	&	0.73	&	0.80	\\
\hline							
\end{tabular}
\end{center}
\caption{Normalized main parameters in the case of the maximum densification ($\gamma=10$) for 3 values of $k$ as shown in Fig. \ref{fig:gam5 vs 3k}. 
$\psi(R_0)/\psi(0)$ is the fractional intensity at the aperture edge compared to that at the center, $I_0$ is the on-axis irradiance, $E_0$ is the energy in the central lobe, $E_0/P_{tot}$ corresponds to the fraction of the energy contained in the central lobe on the total energy, $I_1/I_0$ represents the ratio of the irradiance of the maximum side-lobe to the central peak, $P_{transm}$ is the fractional transmitted power.}
\label{tab:densif vs k}
\end{table}

Neither the geometry of the array (function of number, arrangement and shape of the sub-pupils), nor the densification factor has an influence on the optimum $k$ value. To properly adjust the diaphragm diameter $Do$ with the gaussian field
distribution of each sub-pupil, the output focal length of the
fiber pupil densifier is chosen in such a way that

\begin{equation}
F_o=0.36 \frac{\lambda}{\omega_0}\,R_i\,\gamma
\end{equation}

$R_o$ is fixed by the $\gamma$ value, such as $R_o=\gamma\, R_i$. $F_o$ is imposed by the wavelength and by the characteristics of the fibers, and is also proportional to $\gamma$ to provide the same angular aperture of the exit beams.

\begin{table}
\begin{center}
\begin{tabular}{lccc}
\hline							
$\gamma$	&	1	&	5	&	10	\\
\hline							
$\psi(R_0)/\psi(0)$	&	0.38	&	0.31	&	0.30	\\
$I_0$	&	0.66	&	0.65	&	0.65	\\
$E_0$	&	0.66	&	0.66	&	0.67	\\
$I_1/I_0$	&	1.00	&	1.14	&	1.55	\\
$E_0/P_{tot}$	&	0.89	&	0.90	&	0.92	\\
$P_{transm}$	&	0.75	&	0.73	&	0.73	\\
\hline							
\end{tabular}
\end{center}
\caption{Normalized main parameters in the case of the optimum truncated gaussian illumination ($k=1.1$) for 3 values of $\gamma$. See the definitions in Tab. \ref{tab:densif vs k}.}
\label{tab:gauss vs uniform}
\end{table}

Table \ref{tab:gauss vs uniform} compares the case of the truncated gaussian beams to a uniform illumination. The on-axis irradiance is approximately $35$ per cent of what would be obtained for uniform illumination of the full aperture. The maximum level of the secondary maxima $I_1/I_0$ increases drastically with $\gamma$ whereas the fraction of the total energy contained in the central lobe $E_0/P_{tot}$ does not decrease much. It means that most of the light spreading from the central lobe falls into the few side-peaks next to the center, according to the densification effect which reduces the number of peaks in the image.

Thus, the gaussian field of the fibers increases the spread effect of light in the focal-plane by an energy dilution from the central peak to the side-lobes. This effect should be treated by specific deconvolution algorithms for imaging applications. It must be taken into account for very high dynamic range applications, such as coronagraphy. The use of pupil apodisation techniques \citep{aim05} or of beam shaping techniques \citep{hof00} may also improve the quality of the imaging process.

A direct imager called VIDA \citep{lar05} has been proposed as a second generation instrument for the VLTI. It consists of a densified pupil beam combiner using single-mode fibers in optical wavelength, associated with a coronagraphic mode mounted at the densified focus. A demonstrator is being developped in laboratory \citep{pat06} in order to prepare the feasibility studies of VIDA and to evaluate the technological requirements for future projects of hypertelescopes.

\label{lastpage}

\end{document}